\documentclass[10pt,twoside,preprint]{elsarticle}

\usepackage{amssymb}
\usepackage{amsmath}
\usepackage{subfigure}
\usepackage{graphicx}
\usepackage{caption2}

\usepackage{algorithm} 
\usepackage{algorithmic} 
\usepackage{multirow} 
\usepackage{amsmath}
\usepackage{xcolor}
\usepackage{float}

\usepackage{amssymb}
\usepackage{amsthm}
\newtheorem{theorem}{Theorem}[section]

\theoremstyle{definition}
\newtheorem{definition}[theorem]{Definition}
\theoremstyle{remark}

\usepackage{geometry}   
\usepackage[]{caption2}

\begin{document}
\begin{frontmatter}

\title{Integrated structure investigation in complex networks by label propagation}


\author[rvt,focal,oth]{Tao Wu\corref{cor1}}
\ead{wutaoadeny@gmail.com}
\author[rvt]{Yuxiao Guo}
\author[rvt,focal,oth]{Leiting Chen}
\author[els]{Yanbing Liu}

\cortext[cor1]{Corresponding author}


\address[rvt]{School of Computer Science and Engineering, University of Electronic Science and Technology of China, Chengdu 611731, China}
\address[focal]{Institute of Electronic and Information Engineering in Dongguan, University of Electronic Science and Technology of China, Chengdu 611731, China}
\address[els]{Chongqing University of Posts and Telecommunications, Chongqing 400065, China}
\address[oth]{Digital Media Technology Key Laboratory of Sichuan Province, Chengdu 611731, China}

\begin{abstract}
The investigation of network structure has important significance to
understand the functions of various complex networks. The
communities with hierarchical and overlapping structures and the
special nodes like hubs and outliers are all common structure
features to the networks. Network structure investigation has
attracted considerable research effort recently. However, existing
studies have only partially explored the structure features. In this
paper, a label propagation based integrated network structure
investigation algorithm (LINSIA) is proposed. The main novelty here
is that LINSIA can uncover hierarchical and overlapping communities,
as well as hubs and outliers. Moreover, LINSIA can provide insight
into the label propagation mechanism and propose a parameter-free
solution that requires no prior knowledge. In addition, LINSIA can
give out a soft-partitioning result and depict the degree of
overlapping nodes belonging to each relevant community. The proposed
algorithm is validated on various synthetic and real-world networks.
Experimental results demonstrate that the algorithm outperforms
several state-of-the-art methods.
\end{abstract}


\end{frontmatter}

\section{Introduction}\indent

Complex networks provide a way to represent complex systems of
interacting objects, where nodes denote the objects and edges
describe the interactions between them. For example, in ecological
systems, nodes represent lives and edges represent dependencies, and
in protein association networks, nodes represent proteins and edges
represent physical interactions. Common sense dictates that the
network structure is characterized by at least three common
features: communities, hubs and outliers. Network communities are
often correspond to groups of nodes that share a common property,
role or function. The nodes within one community are more likely to
interact to each other than to the rest of the network. Community
structure allows for understanding the network functions that cannot
be studied when considering only individual object or the entire
system. Moreover, hubs are the nodes that bridge multiple
communities, and hubs identification is crucial to network
exploration. For example, hubs in epidemiology network could be the
nodes that spread epidemics across groups, and immunizing the hubs
could help to prevent the spread of epidemics. In addition, the
identification of outliers that are marginally connected with the
community members is also crucial to networks, as they can be used
for abnormal nodes detection in complex networks. The network
structure investigation is an important aspect of complex networks
research.

In the issue of network structure investigation, early work mainly
focused on the community detection in naive networks
(non-hierarchical and non-overlapping structures). Many classical
methods have been proposed to detect the community structure of
complex networks, including division methods [1,2], agglomerative
methods [3,4], matrix related methods (non-negative matrix
factorization [5], spectral method [6]), model-based methods (label
propagation methods [7,8], mixture models [9], stochastic block
models [10,11]), etc. In a word, the traditional detection problem
about the community structure in naive networks is likely to be
properly solved in the last decade.

However, the problem of community detection becomes much harder for
the networks with complex communities where hierarchical and
overlapping structures emerge at the same time and tangle with each
other. In real-world networks, communities are always nested, and
the networks exhibit hierarchical community structure, such as the
organization of a large company. Moreover, to use a social metaphor,
common sense goes that people can belong to different social
communities, depending on their friends, professions, hobbies, etc.
In network terms, each node can be shared between communities,
forming overlapping communities. Thus, the networks demand methods
that are able to detect complex community with hierarchical and
overlapping structures. So later work paid close attention to the
community detection in hierarchical and overlapping cases, which
generally includes two main genres: hierarchical community detection
[12-14] and overlapping community detection [15-22], although
several researchers made a pioneering attempt on hierarchical and
overlapping community detection [23-25].

Based on the above discussion, most of the existing methods detect
community by considering the hierarchical and overlapping structures
separately, and there are few integrated algorithms for hierarchical
and overlapping community detection. Meanwhile, in the issue of
network structure investigation, there are few methods focusing on
hubs and outliers identification, although hubs and outliers are
important, common features of complex networks. Thus, network
structure investigation should also take hubs and outliers
identification under consideration. Therefore, how to discover the
hierarchical and overlapping communities, hubs and outliers
comprehensively and efficiently remains a important task to date.

To better capture the network structures, an intuitive idea is that
one should analyze networks in a local view. Thus the computation of
the community memberships and the roles of each node can be based
only on local information, which can avoid the complex
organizational structures with a variety of properties. In this
paper, we develop a label propagation based integrated network
structure investigation algorithm (LINSIA). The basic idea behind
LINSIA is that there has a network topology related equilibrium
state between the nodes' community label choices, and the reasonable
structure division can be derived based on the label equilibrium
state. LINSIA controls the label propagation process to discover
multi-scale communities at different aggregation levels, and a
community on a large scale corresponds to multiple underlying
communities. The main contributions include: (1) LINSIA can reveal
hierarchical and overlapping communities, as well as hubs and
outliers; (2) LINSIA can provide insight into the label propagation
mechanism and propose a parameter-free solution that requires no
prior knowledge; (3) LINSIA can give out a soft-partitioning result
and depict the degree of overlapping nodes belonging to each
relevant community. Experimental results demonstrate that the
algorithm outperforms several state-of-the-art methods.

The rest of the paper is organized as follows. Section 2 introduces
some influence measures. Section 3 proposes a new label propagation
strategy. Section 4 presents the LINSIA algorithm. Our algorithm is
tested on a diverse set of networks in Section 5. Discussion and
Conclusion are given in Section 6.

\section{ENCoreness based influence measures}\indent

Before we proceed, it is worthwhile to introduce some necessary
definitions for LINSIA. The node influence represents node
importance in the full network, and the label influence measures the
popularity of each community label in label voting process. The node
influence and the label influence are all regulated by an adaptive
variable $\alpha $. In order to compute node importance accurately,
node influence should consider node's global role and its local
topology information comprehensively. Thus, it exploits extended
neighborhood coreness (ENCoreness) centrality [26] to measure node's
global importance and includes node degree to capture node's local
topology.

\begin{definition} [Node Influence]
Let $NI(i)$ indicates the influence of node $i$, and it is defined
as
\begin{equation}
{NI(i) = ENCoreness(i) + \sum\limits_{j \in N(i)} {{{ENCoreness(j)}
\over {{{degree(j)}^\alpha }}}} *{w_{i,j}}}
\end{equation}
where $ENCoreness(i)$ is the ENCoreness centrality of node $i$,
$degree(j)$ is the degree of node $j$, ${w_{i,j}}$ is the weight of
edge ${e_{i,j}}$, $N(i)$ is the neighbor set of node $i$, and
$\alpha $ is an adaptive real variable. The term $ 1 \over
degree(j)^\alpha $ is called normalization factor, which is used to
consider the different influences of nodes with diverse degrees.

\end{definition}

For the purpose of label selection, we define node $i$'s candidate
label set ${LC_i}$ as the union of neighbor nodes' label sets and
let $Lse{t_j}$ denote the label set of node $j$. Then, we let
neighbors vote for each community label in the candidate label set
${LC_i}$ and use the voting support rate as the label influence. The
detailed definition is presented as follows.

\begin{definition} [Label Influence]
Let $LIse{t_i}$ is the label influence set of the labels in the
candidate label set ${LC_i}$, and it is defined as
\begin{equation}
{LIse{t_i} = \{ LI_i^p | LI_i^p = \sum\limits_{j \in N(i){,_{}}q \in
Lse{t_j}} {\delta (p,q)*{{NI(j)} \over {{{degree(j)}^\alpha }}}}
*{{LI_j^q} \over {\sum\limits_{m \in Lse{t_j}} {LI_j^m}
}}*{w_{i,j}}, p \in {LC_i}\} }
\end{equation}
where $LI_i^p$ is the influence of label $p$ on node $i$, function
$\delta (p,q)$ takes value 1 if $p$ is identical to $q$ and 0
otherwise. Other notations mean the same thing as the notations in
Eq. (1). The term ${LI_j^q} \over \sum\limits_{  m \in Lse{t_j}  } {
LI_j^m }$ is the label influence ratio of label $q$ to the all
labels of node $j$, which denotes the degree of node $j$ belonging
to the community with label $q$. For the iterative updating of the
label influence, all nodes are assigned a unique label initially,
and they have the same initial label influence.
\end{definition}

Why do LINSIA needs the variable $\alpha $? Variable $\alpha $
balances the cohesion inside the community and the competition
between communities, and it is related to the equilibrium state
between nodes' community label choices. If setting a small value to
variable $\alpha $, the node influence and label influence of the
core nodes will be much greater than that of the periphery nodes
(because the ENCoreness values of the core nodes are always bigger
than that of the periphery nodes). In this way, the nodes connected
with a few core nodes and many periphery nodes will have same
community labels as the core nodes rather than the periphery nodes.
The way things are going, perhaps all nodes tend to a common label
choice. Reversely, if the value of variable $\alpha $ is large, all
nodes may tend to a unique label choice. So variable $\alpha $
should be determined based on the network topology by self-adaption.

In LINSIA, we assume that $\alpha $ equals to 1.0 initially, and we
compute the total sum of node influence firstly. Then we get the
cumulative sum of node influence sequentially in descending order of
k-shell value [27] until the cumulative sum equals to half of the
total. The ratio $r$ of the number of nodes corresponding to the
cumulative sum to the total number of nodes can be viewed as an
indicator of imbalance degree of influence distributed among network
nodes. To the networks with completely balanced influence
distribution, the ratio $r$ takes value of 0.5. To the networks with
extremely imbalanced influence distribution, we assume that at least
top ten percent of nodes are needed to have half of the total
influence. That is say, we assume the ratio $r$ takes value of 0.1
in extremely imbalanced networks. So the median value of the ratio
is 0.3. Therefore, the value of the variable $\alpha $ is depend on
a function $F(0.3,{\rm{ }}1.0,{\rm{ }}r)$. Thus the variable $\alpha
$ is self-adaptive and generally takes values over an interval of
real number $(0.4, 1.6)$ in networks with different influence
distribution. Then LINSIA computes the normalize factor value and
regulates the label selection behavior to make the label updating
converge to the equilibrium state. The detailed definition is
presented as follows.

\begin{definition} [Adaptive Variable]
The adaptive variable $\alpha $ is defined as
\begin{equation}
{\alpha = 1.0+(0.3-\frac{N^{^{'}}}{N})^{1/3}}
\end{equation}
where $N$ is the total number of network nodes, and ${N}^{'}$ is the
number of nodes conresponding to the cumulative sum in the above
discussion.
\end{definition}

\section{Label propagation strategy}\indent

\subsection{Label selection and hubs}
According to Eq. (2), in non-overlapping networks, we select the
label with the greatest label influence in $LIse{t_i}$ as node $i$'s
community label. In overlapping networks, we amplify each label
influence in $LIse{t_i}$ of node $i$, and select the label whose
influence has the same order of magnitude with the greatest
influence in $LIse{t_i}$ as one of its labels, defined by Eq. (4).
\begin{equation}
{{Lse{t_i}} = \{ p|\log {(\max (LIse{t_i}))^3} - 1 < \log
{(LI_i^p)^3} < \log {(\max (LIse{t_i}))^3}, for  LI_i^p \in
LIse{t_i}\}}
\end{equation}

If node $i$'s label set $Lse{t_i}$ has more than one label, we
define the node as hub node. The formal definition of hubs is
presented as follows
\begin{equation}
Hubs = \{ i|len(Lset_i ) \ge 1, for\;i \in V\}
\end{equation}
where $len(Lset_i)$ is the number of the labels of node $i$, and $V$
is the node set of the network.

\subsection{Label propagation strategy and outliers}
As nodes' label updating depend more on the core nodes' community
labels than that of the periphery nodes, accurate label selection of
the core nodes is crucial to efficient network structure analysis,
which requires updating node labels in ascending order of node
influence. Otherwise, inaccurate labels of the core nodes will
propagate to the full network and disturb the other nodes' label
selection. Moreover, outliers are the nodes that have negligible
connection with communities, and they might not participate in any
community. Namely, only the nodes with small node degree and
independent community label are meaningful to outliers detection.
According to the common sense, the nodes with these features are
generally the periphery nodes. Therefore, the labels of the
periphery nodes with low node influence should be updated
preferentially to strive for the independent community label and the
equilibrium state between the label choices of the core nodes and
the periphery nodes. Thus, we sort nodes according to their node
influence values and update labels in a particular order, i.e., from
a low influence to a high influence.

Based on the above analysis, we can find that outliers detection
will benefit greatly from updating labels in ascending order. The
formal definition of outliers is presented as follows
\begin{equation}
Outliers = \{ i|len(Lset_i ) =  = 1\;\& \;degree(i) \le 2\;\& \;\phi
(label_i ,\;Inlset_i ) =  = 1,\;for\;i \in V\}
\end{equation}
where $len(Lset_i)$ is the number of the labels of node $i$,
$label_i$ is the community label of node $i$, $Inlset_i$ is the set
of initial labels of node $i$'s neighbors and its own, and $\phi
(label_i ,\;Inlset_i )$ equals to 1, if $label_i$ is identical to
 one of the labels in $Inlset_i$.

\subsection{Participation intensity}
In overlapping networks, to represent the correlation degree of
overlapping node to each community, the community detection result
and the label influence are used to define participation intensity
set
\begin{equation}
{PIntensitySe{t_i} = \{ PI_i^q|PI_i^q = {{LI_i^q} \over
{\sum\limits_{m \in Lse{t_i}} {LI_i^m} }}, q \in Lse{t_i}\} }
\end{equation}
where $PI_i^q$ is the intensity of node $i$ participating in the
community with label $q$.

\section{Description of the algorithm}\indent
\subsection{Algorithm LINSIA}
In real-world networks, communities at one level are always nested
into communities at an upper level, and the networks exhibit
hierarchical community structure. Based on the label selection
mechanism and the label propagation strategy defined in section 3,
LINSIA constructs a weighted bottom-up super-network structure to
detect the hierarchical relationship between the communities. The
first step is to detect communities in primitive network and build a
weighted super-network by replacing the communities with
super-nodes. If the community information is known, it builds a
weighted super-network based on the communities directly. The second
step is to find non-overlapping communities in the super-network,
and mapping the communities to the primitive network to merge the
communities detected in the first step. LINSIA executes the last two
steps iteratively until the community structure no longer changes.
The construction process is illustrated by Fig. 1(a). In the
super-network, the edge's weight is defined by
\begin{equation}
w_{i,j}  = \sum\limits_{m \in com_i ,n \in com_j } {\frac{{{{a_{m,n}
} \mathord{\left/
 {\vphantom {{a_{m,n} } {len(C_m ) \cdot len(C_n )}}} \right.
 \kern-\nulldelimiterspace} {len(C_m ) \cdot len(C_n )}}}}{{N_{i,j} }}}
\end{equation}
where $co{m_i}$ is the node set of the community corresponding to
the super-node $i$, ${a_{m,n}}$ is the element of the primitive
network's adjacency matrix, $len(C_m )$ is the number of communities
to which node $m$ related, and $N_{i,j}$ is the number of the nodes
in $co{m_i}$ and $co{m_j}$ that have edge connecting the
communities.

With the super-network structure, LINSIA can find multi-scale
communities at different aggregation levels. Here a large-scale
community at a high level corresponds to multiple small-scale
communities at a a low level, which reflects the hierarchical
relationship between the communities in real-world networks. LINSIA
returns the hierarchical community structure and the best community
division with the greatest modularity Q [4] or extended modularity
EQ [24]. Based on the results, LINSIA gets hubs and outliers
according to Eq. (5) and Eq. (6). The process of LINSIA is
illustrated in Fig. 1 and the pseudo-code is presented in Algorithm
1.

\subsection{Initial label influence and adaptive variable}
In the iterative updating of label influence, all node labels have
the same initial influence. Generally, because the label updating is
based on the label influence ratio, the initialized values would not
influence the label selection result. Fig. 2(a) plots the finding
number of communities with different initial label influence from 1
to 5 on two synthetic networks with different real community number.
From this plot, we can see that LINSIA allows yielding stable
community partitioning with the initial label influence on different
values. So LINSIA is not sensitive to the initial label influence.

Next, we study the influence of the variable $\alpha $ on the number
of communities, and plots the result in Fig. 2(b) with $\alpha $
ranging from 0.5 to 1.5 on a synthetic network with community
information. From this plot, we can see that the Eq. (3) gives a
pretty well estimation to the value of the variable $\alpha $
corresponding to the real community number. So the definition in Eq.
(3) is effective for LINSIA.

\begin{figure}[!h] \small \centering
\includegraphics[width=16.8cm,height=11.6cm]{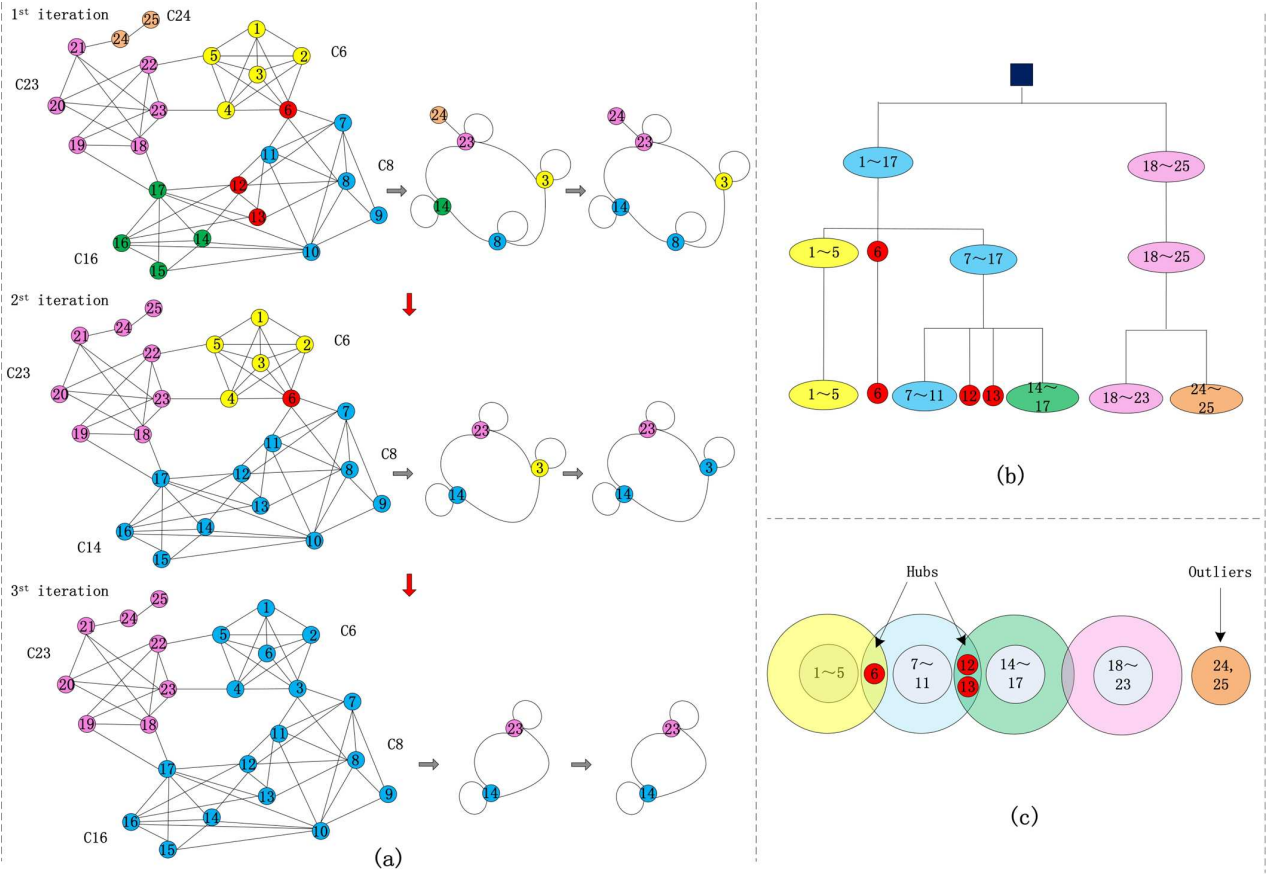}
\caption{The illustration of LINSIA. (a) The construction process of
the bottom-up super-network structure. (b) The hierarchical
community structure of the network. (c) The final network structure
division, which includes 4 communities, 3 hubs and 2
outliers.}\label{fig:1}
\end{figure}

\renewcommand{\algorithmicrequire}{\textbf{Input:}}
\renewcommand{\algorithmicensure}{\textbf{Output:}}
\begin{algorithm}
\caption{LINSIA algorithm}               
\label{alg1}                         
\begin{algorithmic}[1]
\REQUIRE Network $G = (V,E)$, $V = \{ {v_1},{v_2},...,{v_n}\} $.                  
\ENSURE Community division set $Dic = \{ C{S_1},C{S_2},...\}$, the best community division $C{S_{best}} = \{ {C_1},{C_2},...\} $, participation intensity $PIntensitySe{t_i}$, and the set of hubs and outliers.       
\STATE  $t \leftarrow 0$, and assigning a unique label to each node.
\STATE $t \leftarrow t + 1$, and find initial community division
$C{S_t}$.

(a) Compute $NI(i)$ using (1). Arrange nodes in ascending order of
node influence, and store the nodes in vector $NList$.

(b) For each node ${v_i} \in NList$, update its labels according to
(2) and (4).

(c) Repeat (b) until node labels do not change any more. Then, get
initial community division $C{S_t} = \{ {C_1},{C_2},...\} $
according to node labels and $Dic[t] \leftarrow C{S_t}$.

\STATE Find community division set $Dic = \{ C{S_1},C{S_2},...\}$.

(a) Build weighted super-network $S{N_t}$ based on $C{S_t}$ using
(8), then detect non-overlapping community detection on $S{N_t}$ and
get community division $CC{S_t}$ in super-network.

(b) $t \leftarrow t + 1$, and project $CC{S_t}$ into primitive
network to get the merged community division $C{S_t} = \{
{C_1},{C_2},...\} $, then $Dic[t] \leftarrow C{S_t}$.

(c) Iterate the last two steps until the community division no
longer changes.

\STATE Evaluate the divisions in $Dic = \{ C{S_1},C{S_2},...\} $.
Then copy the division $C{S_{i}}$ with the greatest measure value to
$C{S_{best}}$, and copy the underlying division with the
second-greatest measure value to $C{S_{subc}}$.

\STATE Find the set of hubs and outliers based on $C{S_{best}}$
using (5) and (6), and compute $PIntensitySe{t_i}$ using (7).

\STATE Return $C{S_{best}}$, $C{S_{subc}}$, $PIntensitySe{t_i}$, and
the set of hubs and outliers.

\end{algorithmic}
\end{algorithm}

\begin{figure}[!h]
\centering \subfigure[Initial label influence vs. community
number.]{ \label{fig:side:a}
\includegraphics[width=3.1in]{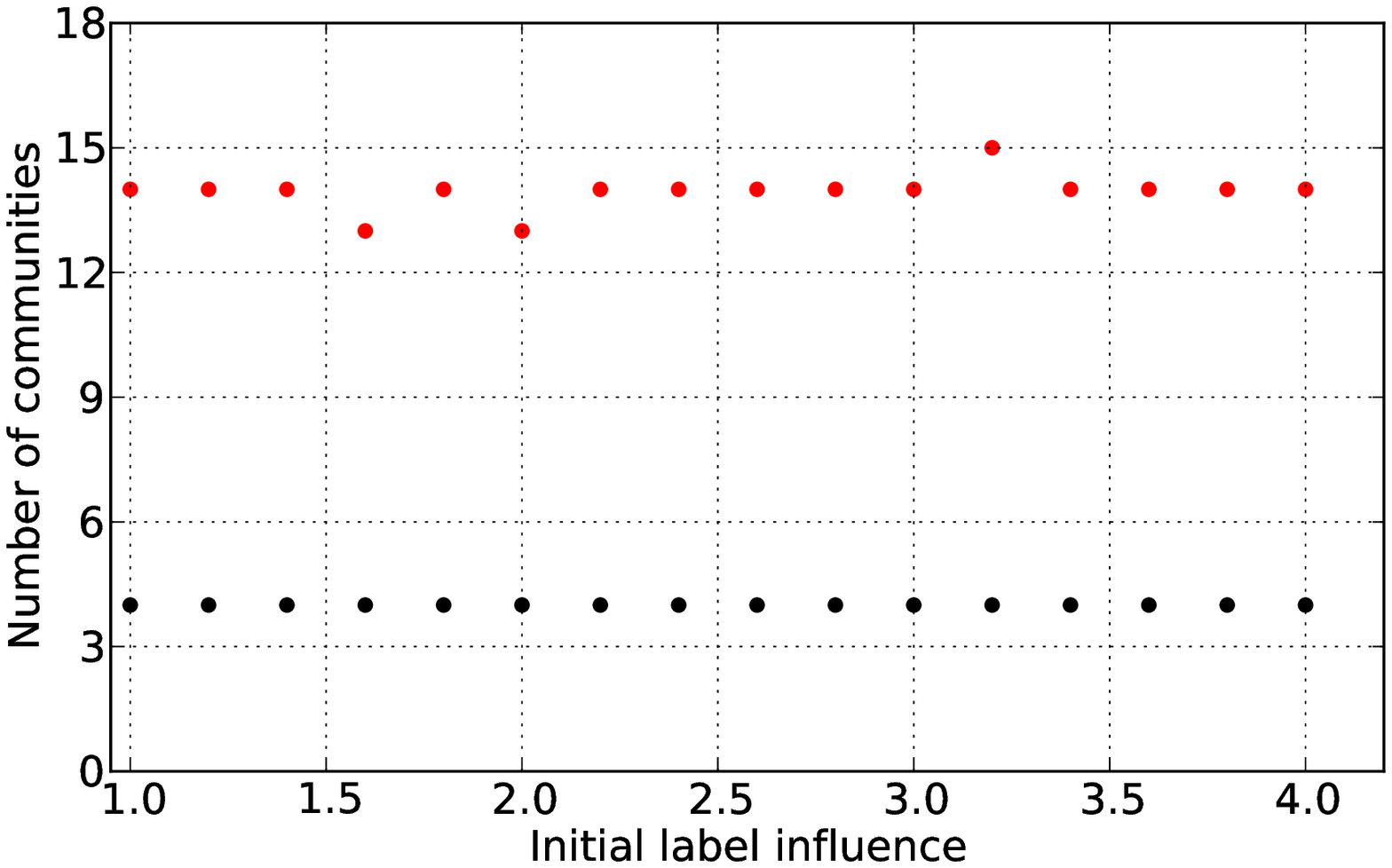}}
\hspace{1ex} \subfigure[Adaptive variable $\alpha$ vs. community
number.]{ \label{fig:side:b}
\includegraphics[width=3.1in]{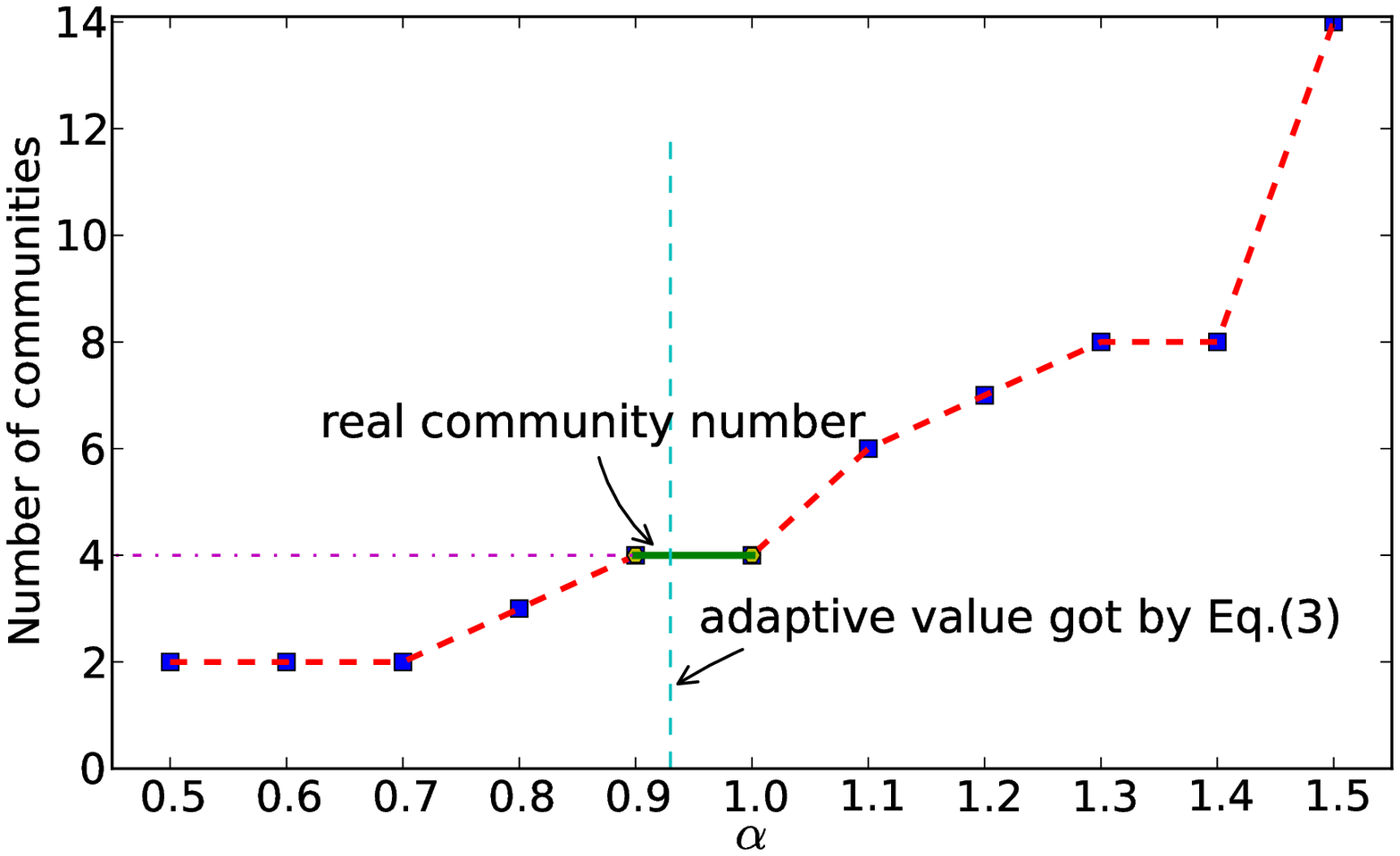}}
\caption{The sensitivity of initial label influence and adaptive
variable $\alpha$ on community detection.}\label{fig:side}
\vspace{\baselineskip}
\end{figure}

\subsection{Computational complexity}
The total complexity of LINSIA is the sum of the complexity of
community detection at all aggregation levels and the complexity of
hubs and outlier identification. The complexity of community
detection at one aggregation level is $O(n) + O(n \cdot log(n)) +
O(t \cdot m)$, where $n$ and $m$ are the number of the nodes and the
edges in the network respectively, $O(n)$ is the complexity of
computing node influence, $O(n \cdot log(n))$ is the complexity of
node ordering, $O(t \cdot m)$ is the complexity of updating
community labels, and $t$ is the number of iterations of label
propagation process. If there are $h$ layers in the bottom-up
super-network structure, the total complexity of community detection
is the sum of the complexities at the aggregation levels
corresponding to the $h$ layers. In addition, the complexity of hubs
and outlier identification is  $ O(n)$. Experiments show that $t$
and $h$ are usually small numbers and the number of nodes and edges
reduces rapidly as the number of the layers increases.

\section{Experimental results}\indent
In this section, to demonstrate the benefits of the proposed
algorithm LINSIA, we apply it on synthetic networks and real-world
networks.

\textbf{Selection of comparison methods.} To evaluate the
performance of LINSIA, we compare it to several representatives of
community detection algorithms.

\textbf{LPA} [7] and \textbf{NIBLPA} [8] are label propagation based
algorithms that use network structure alone as their guide and
require neither optimization of a pre-defined objective function nor
prior information about communities.

\textbf{Newman Fast Algorithm (NF)} [3] is a popular community
detection algorithm, which constructs a ``dendrogram'' by repeatedly
joining communities together in pairs based on the greatest increase
(or smallest decrease) in modularity measure and cuts through the
dendrogram at different levels to give divisions of networks.

\textbf{Louvain} [12] is a well-known modularity based algorithm. It
allows for hierarchical community detection and has low time
complexity.

\textbf{OCSBM} [22] is an overlapping community detection method
based on a principled statistical approach using generative networks
models.

\textbf{EAGLE} [24] deals with a set maximal cliques and adopts an
agglomerative framework. It aims to detect both the hierarchical and
overlapping structures of complex community together.

To evaluate the performance of LINSIA, we compare LINSIA with LPA
and NIBLPA to test its capability of community detection in naive
networks (non-overlapping and non-hierarchical structures). Then we
compare LINSIA with NF and Louvain for hierarchical community
detection, and we compare LINSIA with OCSBM for overlapping
community detection. Finally, we compare LINSIA with EAGLE in the
detection of hierarchical and overlapping communities. For all
experiments, we set the initial label influence to 1.0 for LINSIA as
the default value, and we set $\alpha $ to 0.5 for NIBLPA.

\textbf{Evaluation metrics.} To extensively compare different
community detection algorithms with respect to effectiveness, we
evaluate the division results in two ways.

(a) Networks with class label. If community information is already
known, the normalized mutual information NMI [28] and ENMI [29] are
used to measure the performance of LINSIA for disjoint community
detection and overlapping community detection respectively. NMI is
proposed to provide a indication of the shared information between a
pair of clusterings and ENMI is the extension of NMI for the
similarity quantification of the set of true clusters and the set of
detected clusters by taking overlapping nodes under consideration.

(b) Networks without class label. Since the ground-truths are
unknown, the modularity Q [4] and extended modularity EQ [24] are
used to evaluate the quality of the disjoint communities and the
overlapping communities. Modularity Q is proposed to quantify the
community structure in a network corresponds to a statistically
surprising arrangement of edges and EQ is the extension of Q for the
goodness evaluation of overlapped community decomposition.

\begin{table}[!h]
  \begin{minipage}[b]{0.5\textwidth}
    \centering
    \caption{Parameters of synthetic networks.}
    \includegraphics[width=2.2in]{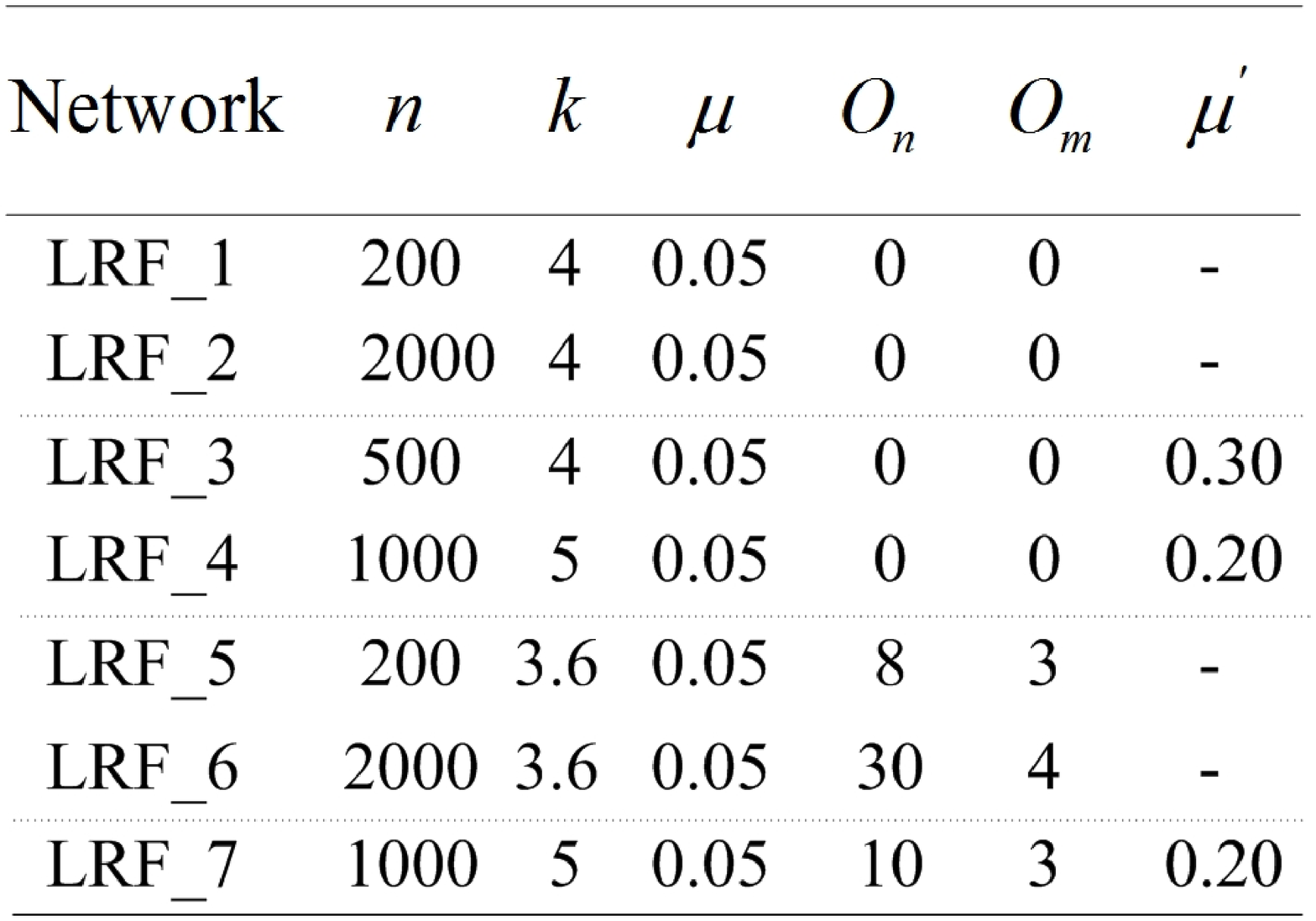}
  \end{minipage}%
  \begin{minipage}[b]{0.5\textwidth}
    \centering
    \caption{Analysis of synthetic networks.}
    \includegraphics[width=3.4in]{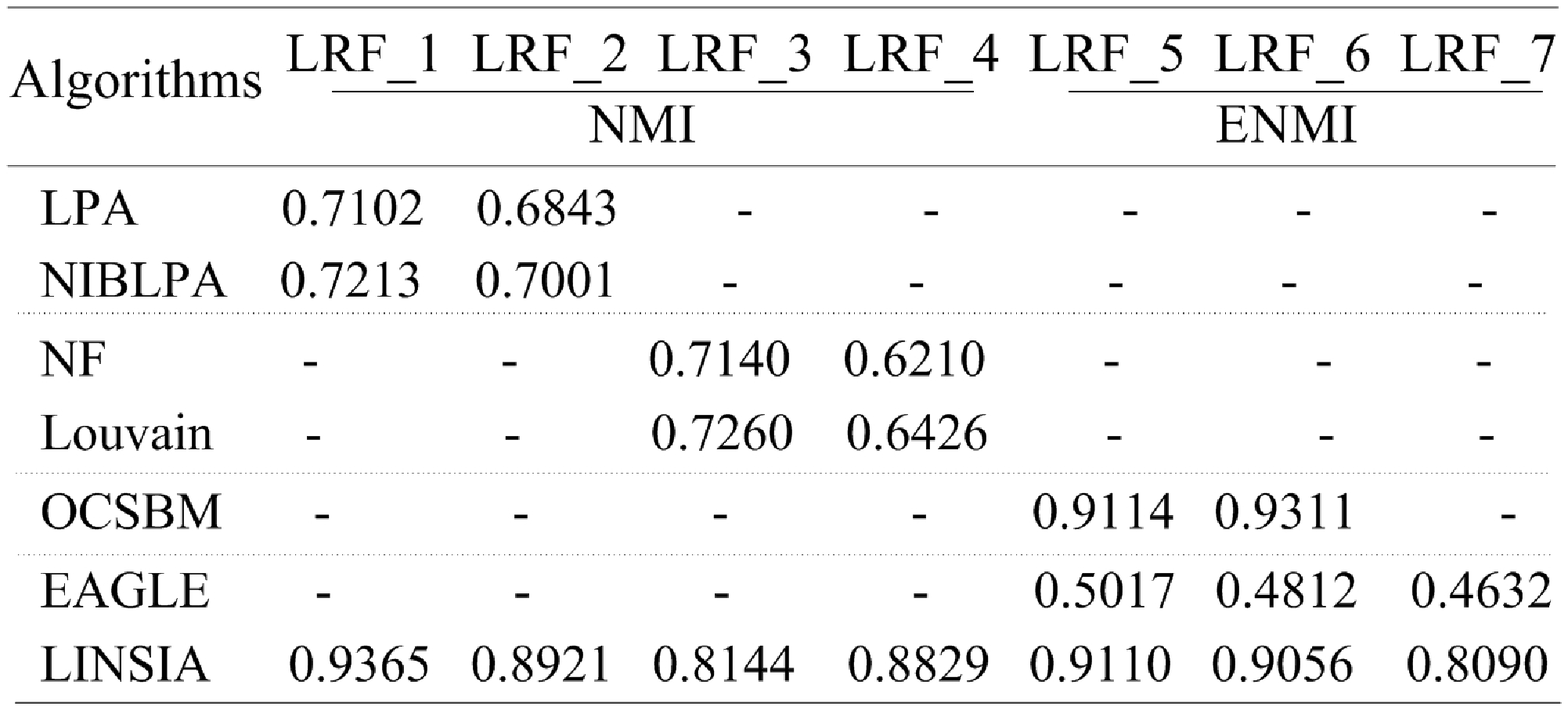}
  \end{minipage}
\end{table}

\makeatletter\def\@captype{table}\makeatother
\begin{minipage}{.45\textwidth}
\centering \caption{Real-world networks.}
\includegraphics[width=2.5in]{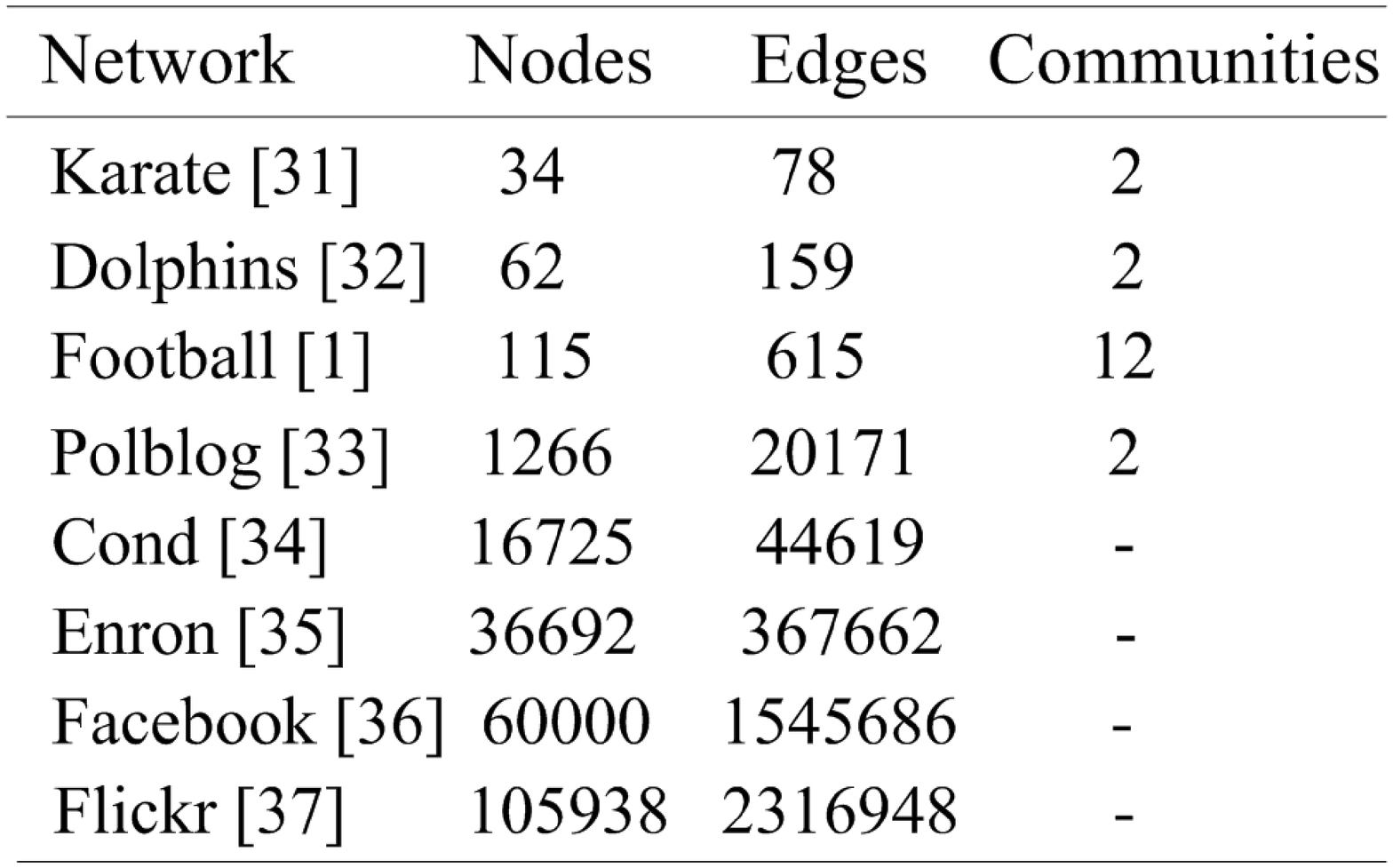}
\end{minipage}
\makeatletter\def\@captype{figure}\makeatother
\begin{minipage}{.45\textwidth}
\centering \includegraphics[width=1.85in]{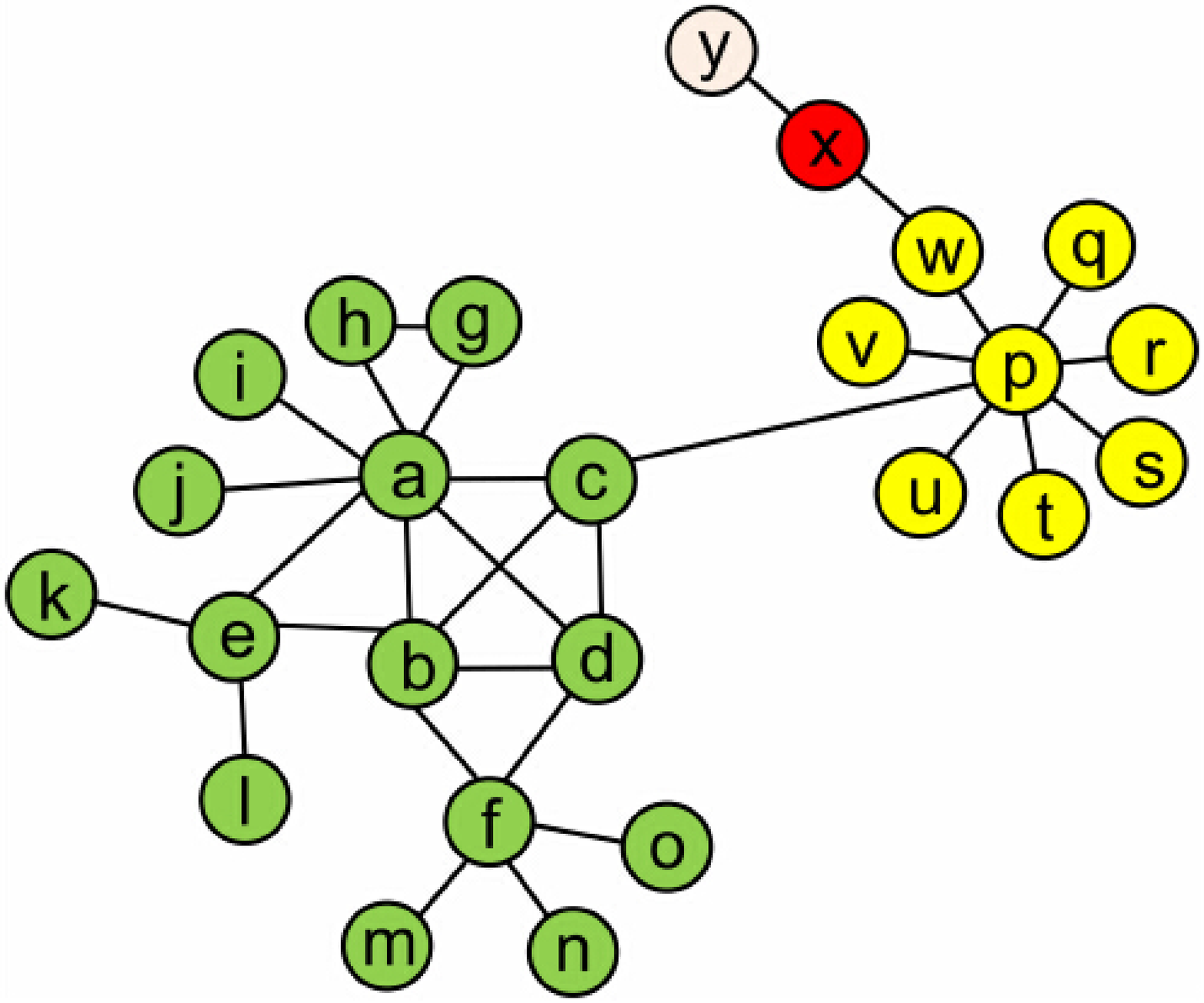} \caption{Hubs and
outliers identification test.}
\end{minipage}

\subsection{Synthetic networks}\indent
In this section, we generate several synthetic networks based on LFR
benchmark networks [30] to evaluate the performance of LINSIA. There
are many parameters to control the generated networks, including the
number of nodes $n$, the average node degree $k$, the mixing ratio
$\mu$ (each vertex shares a fraction $\mu$ of its edges with nodes
in other communities), the number of overlapped nodes ${O_n}$, and
the number of memberships of each overlapped node ${O_m}$. In
addition, parameter $\mu^{'}$ is used to denote the mixing ratio for
the communities at the high level in networks with hierarchical
structures. In our experiments, LRF\_1 and LRF\_2 are naive
synthetic networks, LRF\_3 and LRF\_4 are synthetic networks with
hierarchical structures, LRF\_5 and LRF\_6 are synthetic networks
with overlapping structures, and LRF\_7 is a synthetic network with
overlapping and hierarchical structures. The parameters for the
networks are listed in Table 1.

The results of the methods on the synthetic networks are shown in
Table 2. We can see that LPA and NIBLPA can detect satisfactory
community divisions in the naive network LRF 1 and LRF 2. Meanwhile,
NF and Louvain yield acceptable results for the network LRF 3 and
LRF 4. OCSBM almost achieves the perfect community divisions given
the number of communities on the network LRF 5 and LRF 6. Although
EAGLE allows for hierarchical and overlapping community detection,
its accuracy in the network LRF 5, LRF 6 and LRF 7 is low. By
contrast, LINSIA can find communities which almost exactly match the
ground-truths on all the networks. So LINSIA is competitive with the
other methods in the synthetic networks.

In addition, we design a synthetic network to test the capability of
LINSIA on hubs and outliers identification. The result is shown in
Fig. 3, in which the nodes with the same community label are
rendered by the same color. The result shows that the hub node (in
red color) serves as a bridge between the nodes with different
community label, and the outlier node (in pink color) is marginally
connected with a community (in yellow color). So LINSIA can find
rational hubs and outliers in the synthetic network.

\subsection{Real-world networks}\indent
In this section, we conduct experiments on a wide range of
real-world networks shown in Table 3. In order to sufficiently
evaluate LINSIA's performance, we first test it in terms of
non-overlapping and non-hierarchical community detection,
hierarchical community detection, overlapping community detection
and overlapping and hierarchical community detection. Then we test
it for hubs and outliers identification. Moreover, in addition to
measuring the results with evaluation metrics, we provide the
results of LINSIA on several real-world networks in visual graphics
to demonstrate LINSIA's capability of community, hubs and outliers
detection.

(1) Non-overlapping and non-hierarchical community detection

We compare our algorithm with LPA, NIBLPA for non-overlapping and
non-hierarchical community detection, and Table 4 shows the results
of the methods. We can see that LINSIA gets a better result than LPA
and NIBLPA.

\begin{table}[!h]\small \centering
\caption{The performance of different algorithms for non-overlapping
and non-hierarchical community detection.}
\includegraphics[width=6.8in]{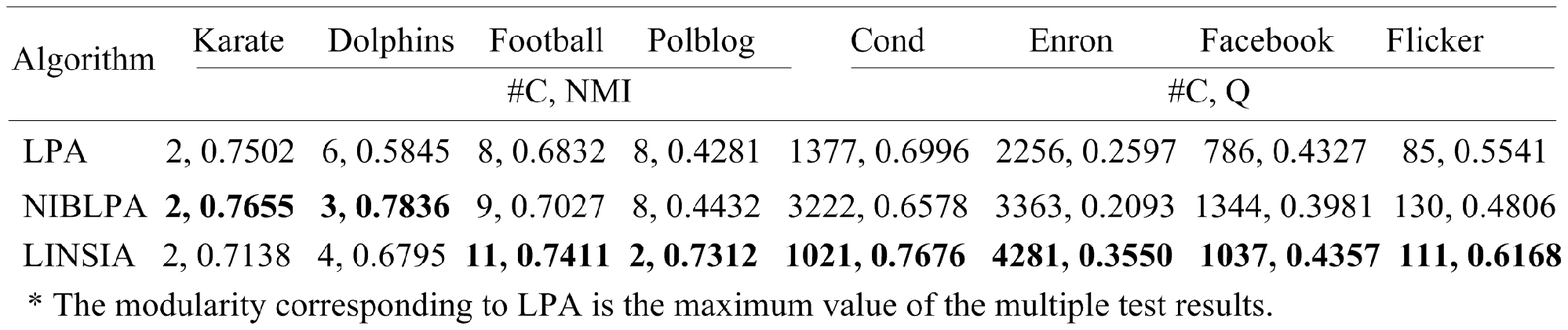}
\end{table}

\begin{figure}[!h]\small \centering
\includegraphics[width=3.3in]{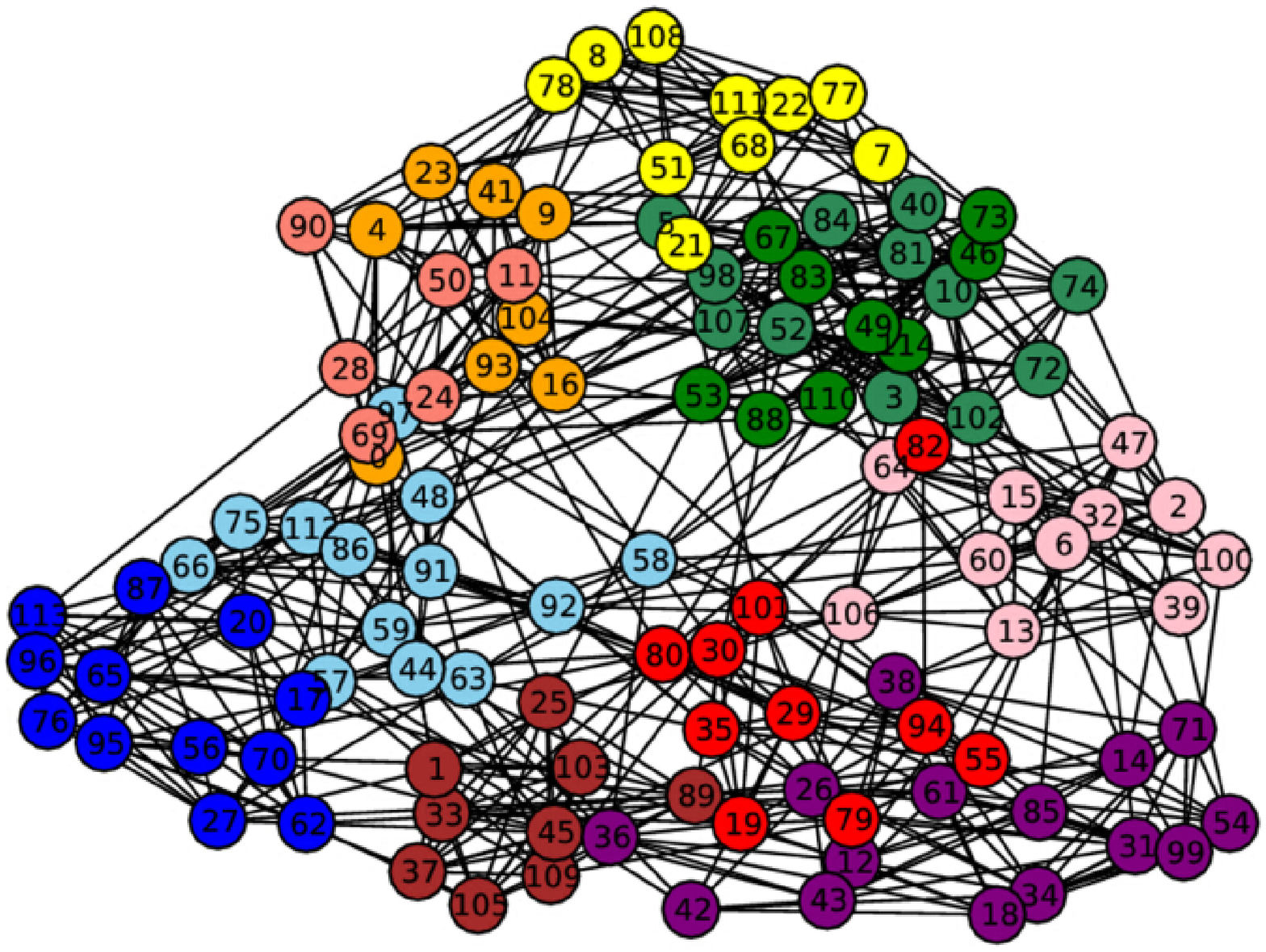}
\caption{Result of LINSIA on Football network.}\label{fig:4}
\end{figure}

\noindent Moreover, we find that LPA is a parameterized method and
it gets different community divisions on the same network, which is
inapplicable in practical problems. It is easy to understand that
LINSIA outperforms LPA and NIBLPA since LINSIA improves the
traditional method in terms of influence definition, label
propagation and label selection. Below we analyze one of the
networks in detail. The American College Football network is the
network of games between Division IA colleges. There are 115 teams
and 12 different conferences except for 8 independent teams. Fig. 4
plots the communities which are detected by LINSIA. It is
interesting to note that LINSIA automatically finds 11 strong sense
communities with NMI 0.7411. From this figure, we can observe that
most of the teams are correctly assigned into corresponding
communities. For LPA and NIBLPA, however, they are difficult to
discover the natural community structure, and the NMI values are
0.6832 and 0.7027 respectively.

(2) Hierarchical community detection

To evaluate the performance of our algorithm on hierarchical
community detection, we compare LINSIA with NF and Louvain, and the
results are shown in Table 5. For the labeled networks, we present
the numbers and the NMI values of the macro communities at the high
level of the hierarchical community structures. For networks without
class information, we compute the average Q of the communities of
the hierarchical community structure. Table 5 shows that LINSIA
achieves a good performance. However, Louvain results in a
relatively low values of NMI, and the performance of NF is a bit
worse. It is not surprise that LINSIA outperforms NF and Louvain
since LINSIA can propagate local labels iteratively, and it can
consider label information about the full network to achieve global
optimal solution in each label selection. By contrast, NF and
Louvain do not have similar mechanism to provide structure
information of the full network for individual decision, which may
leads to locally optimal solution.

\begin{table}[!h]\small \centering
\caption{The performance of different algorithms for hierarchical
community detection.}
\includegraphics[width=6.8in]{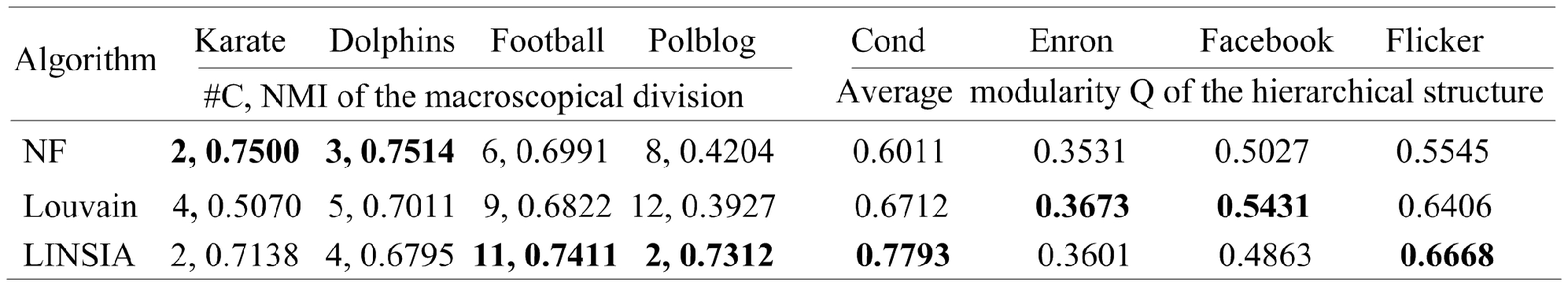}
\end{table}

\begin{table}[!h]\small \centering
\caption{The performance of different algorithms for overlapping
community detection.}
\includegraphics[width=6.8in]{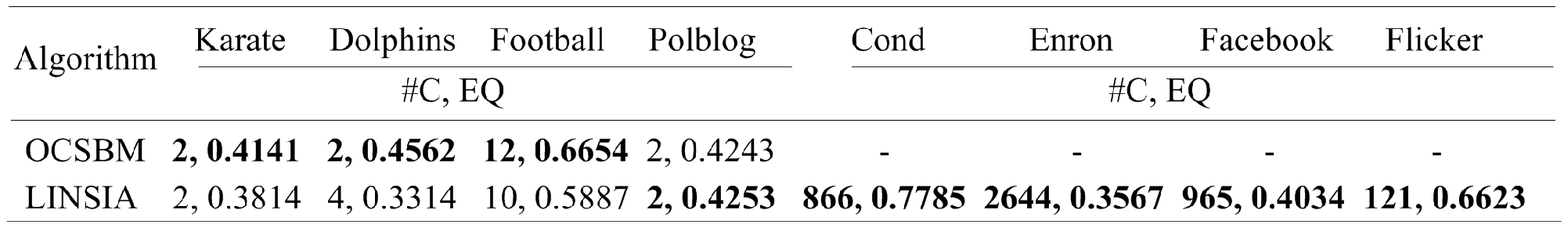}
\end{table}

\begin{figure}[!h]   \small \centering
\includegraphics[width=2.9in]{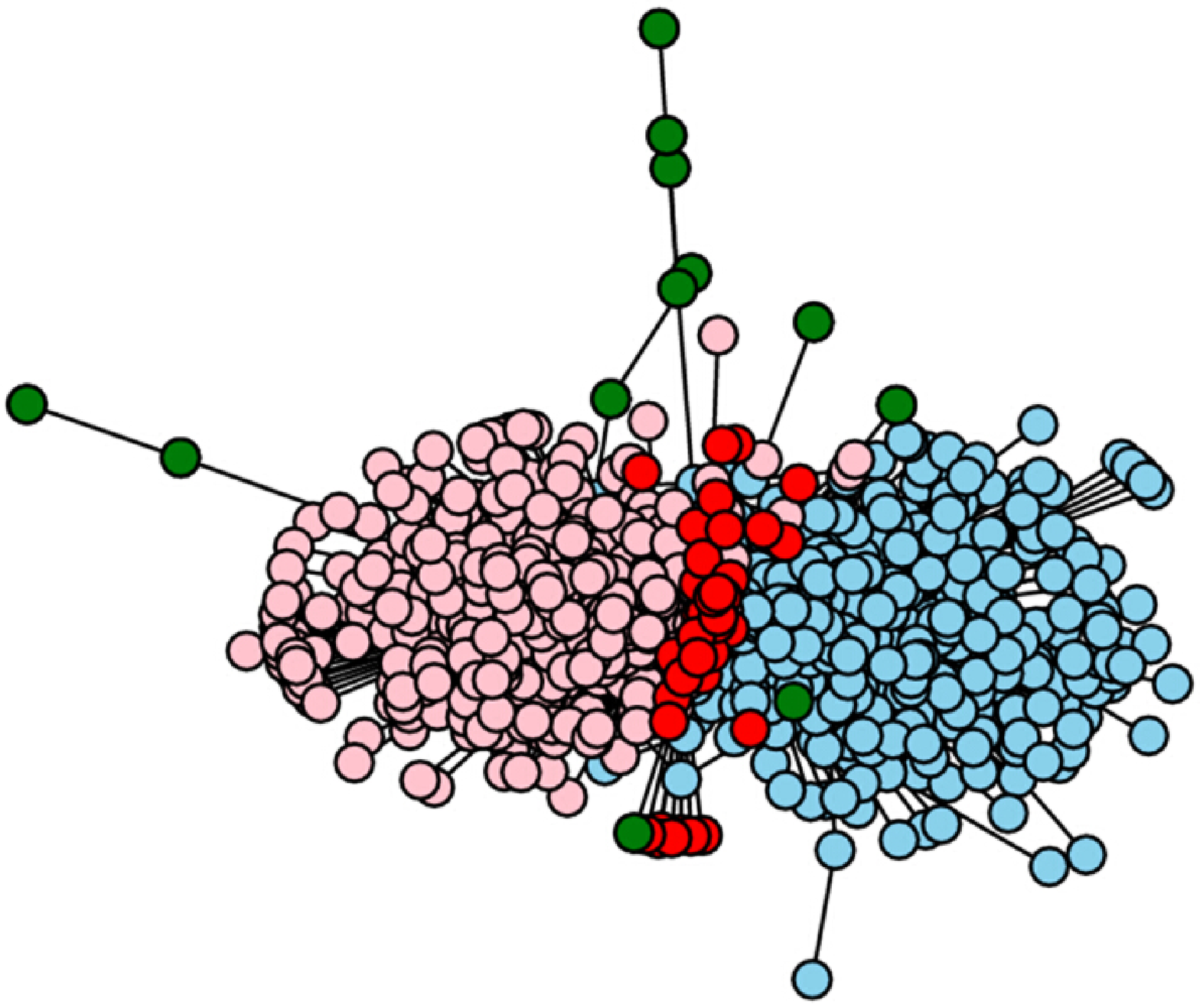}
\caption{Result of LINSIA on Polblog network.}\label{fig:5}
\end{figure}

(3) Overlapping community detection

We compare LINSIA with OCSBM to demonstrate the capability of LINISA
on overlapping community detection, and the results are shown in
Table 6. As there is no overlapping ground-truth information of all
these networks, we use EQ to evaluate the algorithms. Given the
community number, we find that OCSBM can get high-quality community
division in the labeled networks. However, the method is
unsatisfactory in the networks without prior community information.
We can see that LINSIA detects better results than OCSBM in most of
the networks. Below we analyze the Polblog network in detail and
give a graphical display. This network is derived from the links
between weblogs about US politics published around the time of the
2004 presidential election. The links between weblogs were
automatically extracted from a crawl of the front page of the
weblog. Each blog is labeled with `0' or `1' to indicate whether
they are ``liberal'' or ``conservative''. LINSIA groups these
weblogs into two categories, which well represent the corresponding
liberal and conservative weblogs respectively.

(4) Hierarchical and overlapping community detection

As shown in Table 7, we compare LINSIA with EAGLE for hierarchical
and overlapping community detection. As there is no overlapping
ground-truths, we compute the average EQ of the overlapping
communities to measure the quality of the detected hierarchical
community structures of the networks. Meanwhile, we use the number
of the macro communities at the high level of the hierarchical
structures as a evaluation metric. From Table 7, we can see that
LINSIA shows a clear advantage over EAGLE based on the average EQ.
Moreover, the community structure detected by EAGLE tends to have a
large number of communities with small size. The reason is that
EAGLE is a maximum cliques based method. As real-world networks are
generally sparse, most of the maximum cliques detected by EAGLE only
have several nodes.

Dolphins network is an undirected social network representing
frequent associations between 62 dolphins in a community living off
Doubtful Sound, New Zealand, and the hierarchical community
structure detected by LINSIA of the Dolphins network is shown in
Fig. 6. Fig. 6(a) represents the overlapping communities at the high
level, and Fig. 6(b) represents the overlapping communities at the
low level. According to Fig. 6, we can find that the hierarchical
relationship between the overlapping communities. According to [31],
we can conclude that LINSIA can detect almost the real communities
of the network. Furthermore, schematic network is a clique dominated
network which is used for performance evaluation of EAGLE in [24].
For the performance comparison of LINSIA against EAGLE, we apply
LINSIA on the network, and the overlapping communities and the
hierarchical relationship between them are shown in Fig. 7. The
community with blue color in Fig. 7(a) includes the community with
green color and the community with blue color in Fig. 7(b).
According to the network topology, we can conclude that the result
is reasonable. Meanwhile, the community division found by LINSIA is
identical with the result detected by EAGLE in [24]. Thus we can
conclude that EAGLE performs well on clique dominated networks, and
LINSIA can detect rational overlapping and hierarchical communities
on multiple types of networks.

\begin{table}[!h]\small \centering
\caption{The performance of different algorithms for hierarchical
and overlapping community detection.}
\includegraphics[width=6.8in]{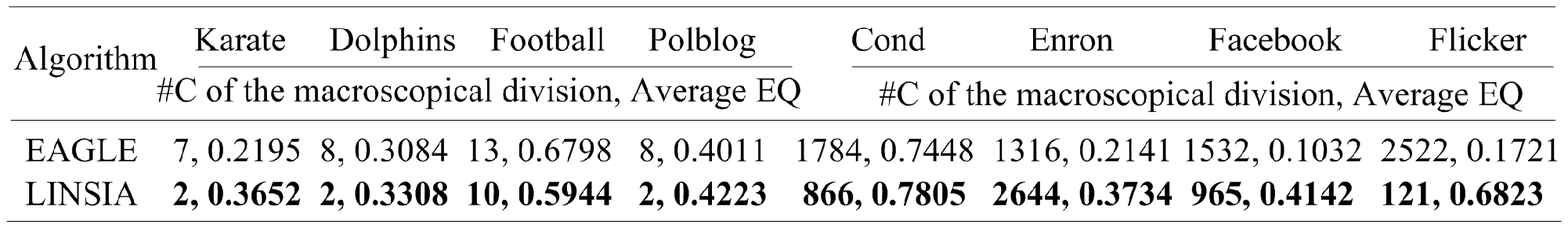}
\end{table}

\begin{figure}[!h]
\centering \subfigure[$EQ$=0.3314, $C$=2.]{ \label{fig:side:a}
\includegraphics[width=3.0in]{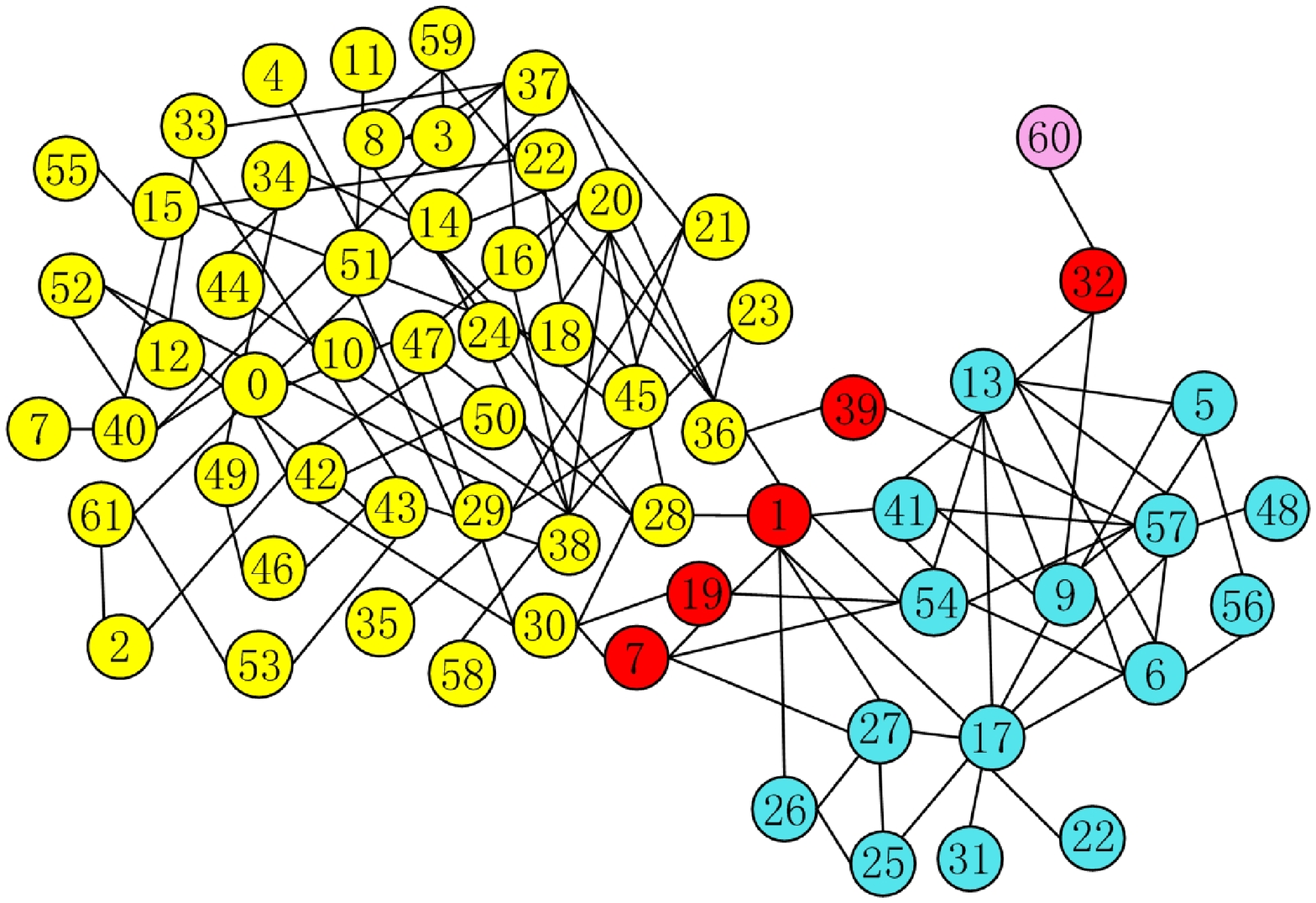}}
\hspace{8ex} \subfigure[$EQ$=0.3301, $C$=3.]{ \label{fig:side:b}
\includegraphics[width=3.0in]{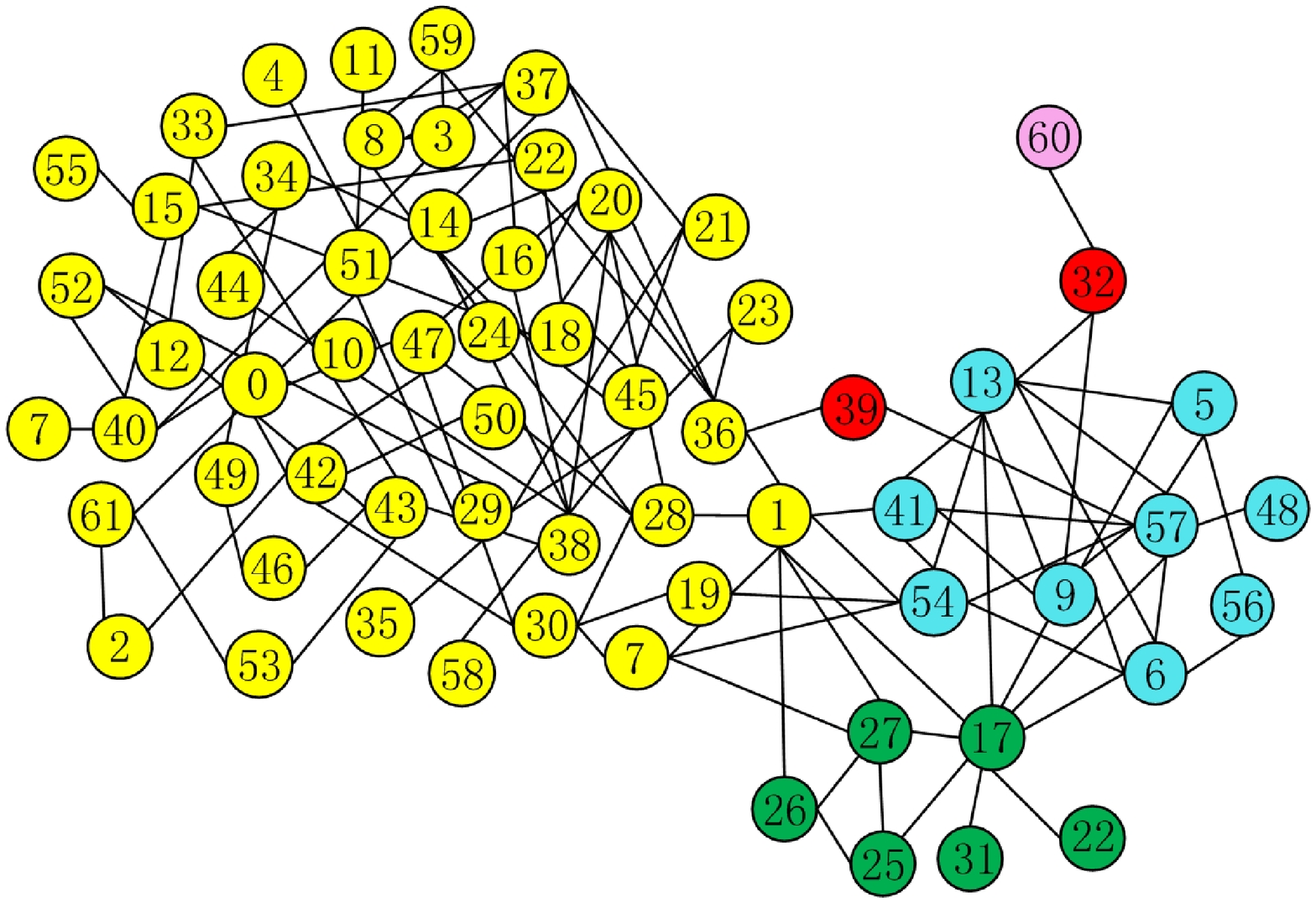}}
\caption{Result of LINSIA on Dolphins network.}\label{fig:side}
\vspace{\baselineskip}
\end{figure}

\begin{figure}[!h]
\centering \subfigure[$EQ$=0.4986, $C$=3.]{ \label{fig:side:a}
\includegraphics[width=1.9in]{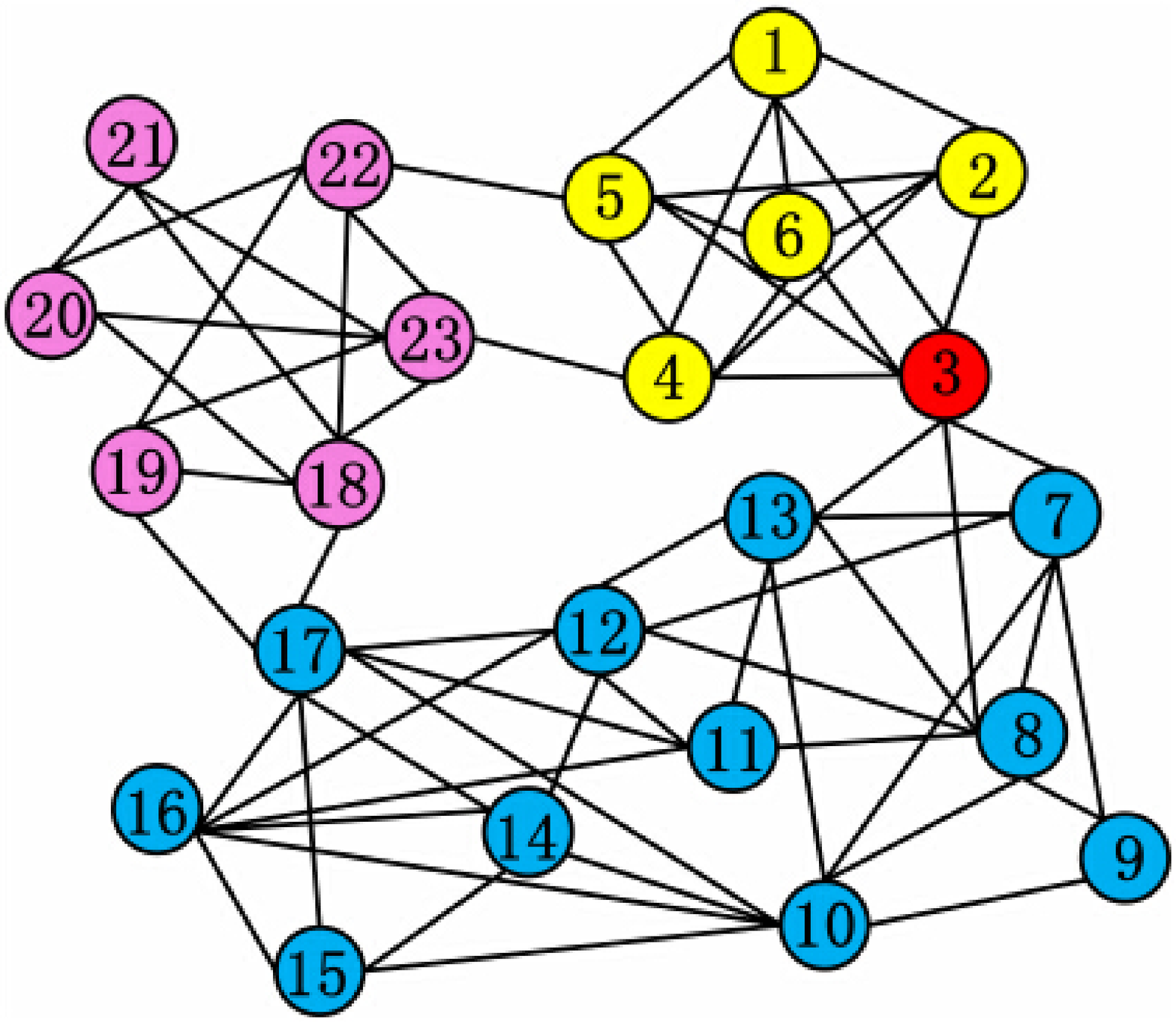}}
\hspace{8ex} \subfigure[$EQ$=0.4801, $C$=4.]{ \label{fig:side:b}
\includegraphics[width=1.9in]{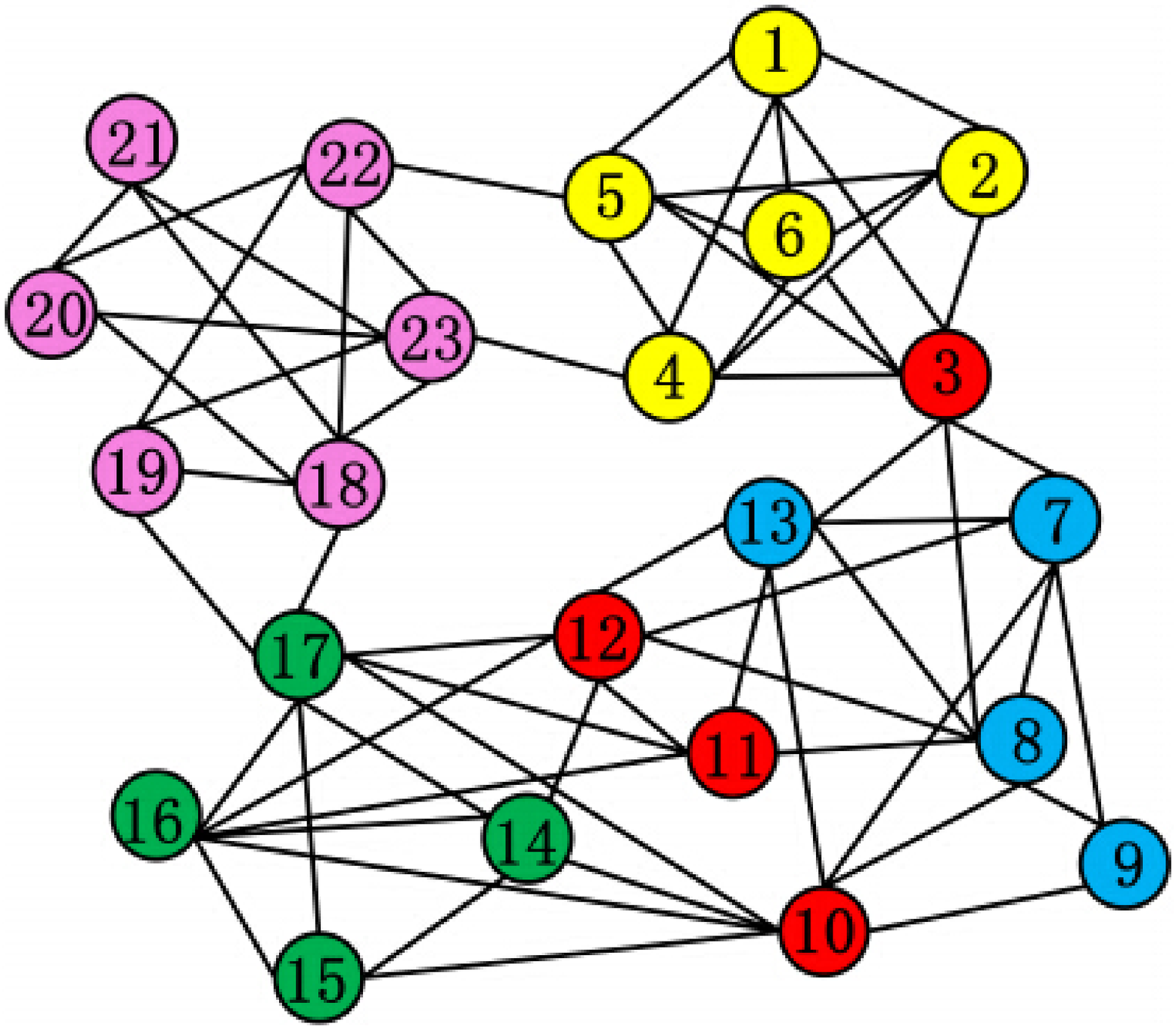}}
\caption{Result of LINSIA on schematic network.}\label{fig:side}
\vspace{\baselineskip}
\end{figure}

(5) Hubs and outliers identification

Given that there is no universal standard for hubs and outliers
evaluation, the capability of LINSIA on hubs and outliers
identification is evaluated based on result analysis.

To evaluate LINSIA's effectiveness on hubs detection, we apply it on
three real-world networks. One of the networks is the Dolphins
network. The hubs of the Dolphins network detected by LINSIA are
shown as red nodes in Fig. 6. Considering the network structure and
community division comprehensively, we can conclude that the
detected hubs are rational. Furthermore, we apply LINSIA on the
Polblog network in Fig. 5 and the schematic network in Fig. 7. We
can conclude that the detected hubs in red color are also
reasonable.

In addition, LINSIA has the advantage of soft-partitioning solution,
and it can depict the degree of hubs belonging to each relevant
community. For example, in the Dolphins network, the participation
intensity set of the hubs in Fig. 6(a) is: \{1:[(57, 0.746), (51,
0.253)], 7: [(57, 0.501), (51, 0.499)], 19: [(57, 0.5708), (51,
0.429)], 32: [(32, 0.600), (57, 0.399)], 39: [(57, 0.428), (51,
0.572)]\}. Take the hub node 39 in the participation intensity set
for example, the item means that node 39 belongs to the community
labeled by 57 (in blue color) with degree 0.428 and belongs to the
community labeled by 51 (in yellow color) with degree 0.572.

LINSIA's capability on outliers identification can also be
demonstrated by the results of the Polblog network in Fig. 5 and the
Dolphins network in Fig. 6. In the results, the outliers in Fig. 5
(in green color) and the outliers in Fig. 6 (in pink color) are
rational according to the network topology. The reason is that our
algorithm proposes a new label updating strategy to make the
periphery nodes have opportunity to compete with the core nodes in
label selection process. For example, in Fig. 6(a), firstly outlier
node 60 selects node 32's label as its community label. Then node 32
acts as an overlapping node after weighting the label influence of
neighbor nodes. In this way, node 60 can exist with its local label
information, instead of being overwhelmed by core nodes' labels. So
our algorithm can find rational outliers efficiently.

In total, the experiments on all real-world networks demonstrate
that LINSIA not only allows for extracting good complex communities
in terms of the internal measures (Q and EQ) and external measures
(NMI and ENMI), but also can identify reasonable hubs and outliers.

\subsection{Running time}\indent

\begin{figure}[!h]   \small \centering
\includegraphics[width=4.5in]{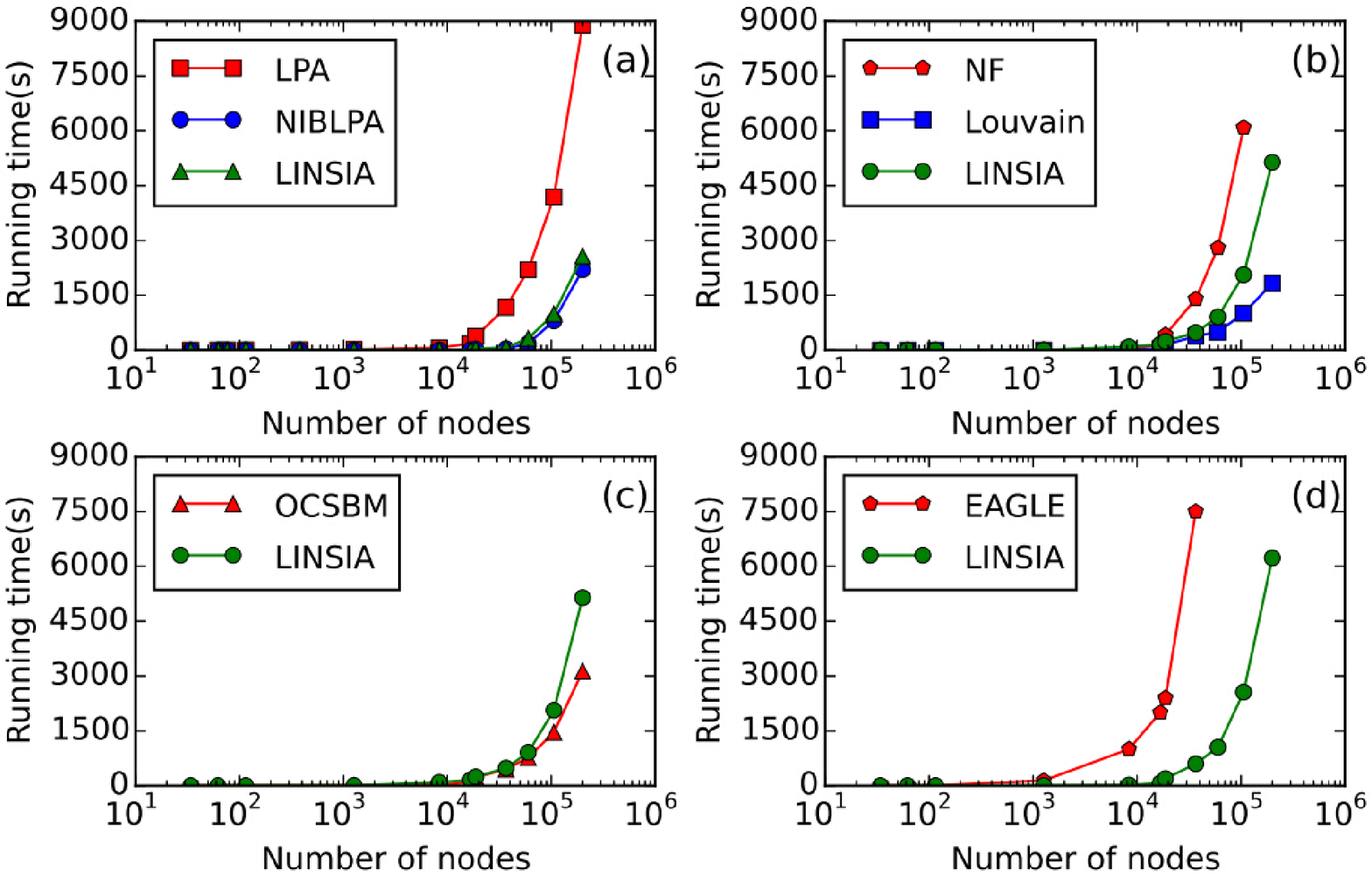}
\caption{Running times comparison. (a) The running times of
algorithms for non-overlapping and non-hierarchical community
detection. (b) The running times of algorithms for hierarchical
community detection. (c) The running times of algorithms for
overlapping community detection. (d) The running times of algorithms
for overlapping and hierarchical community detection.}\label{fig:7}
\end{figure}

To assess and evaluate the scalability of LINSIA with respect to
network size, we compare the running times of LINSIA as well as the
comparison algorithms on the real-world networks listed in Table 3.
Fig. 8 shows the running times of the different community detection
algorithms. We can observe that LINSIA is faster than the other
methods except Louvain and OCSBM. Although Louvain and OCSBM are
faster than LINSIA, Louvain focuses only on hierarchical community
detection, and OCSBM focuses only on overlapping community detection
and requires prior community number for sufficiently good
performance. By contrast, LINSIA is not only applicable to community
detection, but also can  identify hubs and outliers. In addition, we
can find that LINSIA can process the networks with hundreds of
thousands of nodes within two hours. So for the comprehensive
complex network structure analysis, our algorithm LINSIA performs
well in term of running time.

\section{Discussion and Conclusion}\indent

Network structure analysis has always been an important research
topic. Though many related researches have been done, most of them
focus only on one of the structural features. The problem of
investigating network structure comprehensively is very important,
with many potential applications in real-world networks. However,
little has been down in this research direction. In this paper, we
have proposed an integrated network structure investigation
algorithm LINSIA. The proposed algorithm outperforms several
representative network structure analysis algorithms. However, we
are still far away from thoroughly understanding the structure of
complex networks, and some more issues still remain open. In this
paper, the method focuses only on structural aspect. Besides
structure, there are many dynamical processes on complex networks
such as synchronization [38]. We hope the method and results in this
paper can inspire some network structure analysis methods taking the
network dynamics under consideration.

\section*{Acknowledgements}\indent
The work was supported partially by the National Natural Science
Foundation of China (Grant No. 61202255), University-Industry
Cooperation Projects of Guangdong Province (Grant No.
2012A090300001) and the Pre-research Project (Grant No.
51306050102). We thank Tao Zhou and JunMing Shao for their advices.
We also thank Xin Li, Yunpeng Xiao, Xingping Xian and Yuanping Zhang
for enlightening discussions and careful reading of the manuscript.
The authors also wish to thank the anonymous reviewers for their
thorough review and highly appreciate their useful comments and
suggestions.

\section*{References}


\begin{thebibliography}{33}
\small

\bibitem{1}                      
M. Girvan, M. E. J. Newman,     
\textit{``Community structure in social and biological networks"},          
Proc. Natl. Acad. Sci. USA 99 (2002) 7821¨C7826.                   



\bibitem{2}                      
M. Girvan, M. E. J. Newman,                       
\textit{``Finding and evaluating community structure in networks"},  
Phys. Rev. E 69 (2004) 026113.                  




\bibitem{3}                      
M. E. J. Newman,                      
\textit{``Fast algorithm for detecting community structure in networks"},  
Phys. Rev. E 69 (2004) 066133.                  



\bibitem{4}                        
M. E. J. Newman,                              
\textit{``Modularity and community structure in networks"},
Proc. Natl. Acad. Sci. USA 103 (2006) 8577-8582.                



\bibitem{5}                        
J. Yang, J. Leskovec,                               
\textit{``Overlapping community detection at scale: a nonnegative
matrix factorization approach"},
WSDM' 13, pp. 587-596.              




\bibitem{6}                      
M. E. J. Newman,                      
\textit{``Spectral methods for network community
detection and graph partitioning"},  
Phys. Rev. E 88 (2013) 042822.                 




\bibitem{7}                      
U. N. Raghavan, R. Albert, S. Kumara,                     
\textit{``Near linear time algorithm
to detect community structures in large-scale networks"},  
Phys. Rev. E 76 (2007) 036106.                




\bibitem{8}                      
Y. Xing, F.R. Meng, Y. Zhou,                    
\textit{``A Node Influence Based Label
Propagation Algorithm for Community Detection in Networks"},  
The Sci. World J. 2014 (2014) 627581.             




\bibitem{9}                        
M. E. J. Newman, E. Leicht,                              
\textit{``Mixture models and exploratory analysis in networks"},
Proc. Natl. Acad. Sci. USA 104 (2007) 9564-9569.              



\bibitem{10}                        
Brian Karrer, M. E. J. Newman,                             
\textit{``Stochastic blockmodels and community structure in
networks"},
Phys. Rev. E 83 (2011) 016107.              






\bibitem{11}                        
Zh. Q. Xu, Y. P. Ke, Y. Wang,                        
\textit{``A Fast Inference Algorithm for Stochastic Blockmodel"},
ICDM'14, pp. 620-629.             



\bibitem{12}                        
V. D. Blondel, J. L. Guillaume, R. Lambiotte, E. Lefebvre,                             
\textit{``Fast unfolding of communities in large networks"},
J. Stat. Mech. (2008) P10008.             





\bibitem{13}                        
J. Huang, H. Sun, J. Han, H. Deng, Y. Sun,                            
\textit{``SHRINK: A Structural Clustering Algorithm for Detecting
Hierarchical Communities in Networks"},
CIKM'10, pp. 219-228.             


\bibitem{14}                        
M. Sales-Pardo, R. Guimera, A. A. Moreira, L. A. Amaral,                             
\textit{``Extracting the hierarchical organization of complex
systems"},
Proc. Natl. Acad. Sci. USA 104 (2007) 15224-15229.             



\bibitem{15}                      
S. Gregory,                     
\textit{``Finding overlapping communities in networks by label
propagation"},  
New J. Phys. 12 (2010) 103018.               


\bibitem{16}                        
G. Palla, I. Derenyi, I. Farkas, T. Vicsek,                            
\textit{``Uncovering the overlapping community structure of complex
networks in nature and society"},
Nature 435 (2005) 814-818.             



\bibitem{17}                        
J. R. Xie, B. K. Szymanski, X. M. Liu,                        
\textit{``Slpa: Uncovering Overlapping Communities in Social
Networks via a Speaker-Listener Interaction Dynamic Process"},
ICDMW'11, pp. 344-349.     


\bibitem{18}                        
D. Chen, M. Shang, Z. Lv, Y. Fu,                        
\textit{``Detecting overlapping communities of weighted networks via
a local algorithm"},
Physica A 389 (2010) 4177-4187.     



\bibitem{19}                        
Y. Z. Cui, X. Y. Wang, J. Q. Li,                          
\textit{``Detecting overlapping communities in networks using the
maximal sub-graph and the clustering coefficient"},
Phys. A 405 (2014) 85-91.           


\bibitem{20}                        
P. K. Gopalan, D. M. Blei,                             
\textit{``Efficient discovery of overlapping communities in massive
networks"},
Proc. Natl. Acad. Sci. USA 110 (2013) 14534-14539.             



\bibitem{21}                        
B. J. Sun, H. W. Shen, X. Q. Cheng,                          
\textit{``Detecting overlapping communities in massive networks"},
EPL. 108 (2014) 68001.          



\bibitem{22}                        
B. Ball, B. Karrer, M. E. J. Newman,                          
\textit{``Efficient and principled method for detecting communities
in networks"}, Phys. Rev. E 84 (2011) 036103.        




\bibitem{23}                        
A. Lancichinetti, S. Fortunato, J. Kertesz,                          
\textit{``Detecting the overlapping and hierarchical community
structure of complex networks"},
New J. Phys. 11 (2009) 033015.          




\bibitem{24}                        
H. Shen, X. Cheng, K. Cai, M.B. Hu,                        
\textit{``Detect overlapping and hierarchical community structure in
networks"},
Phys. A 388 (2009) 1706-1712.          



\bibitem{25}                        
F. Havemann, J. Glaser, M. Heinz, M. Struck,                        
\textit{``Identifying Overlapping and Hierarchical Thematic
Structures in Networks of Scholarly Papers: A Comparison of Three
Approaches"},
PLoS ONE 7 (2012) e33255.        


\bibitem{26}                        
J. Bae, S. Kim,                        
\textit{``Identifying and ranking influential spreaders in complex
networks by neighborhood coreness"},
Phys. A 395 (2014) 549-559.       



\bibitem{27}                        
M. Kitsak, L. K. Gallos, S. Havlin, F. Liljeros, L. Muchnik, H.
E. Stanley, H. A. Makse,                        
\textit{``Identification of influential spreaders in complex
networks"},
Nature Phys. 6 (2010) 888-893.      



\bibitem{28}                        
A. Strehl, J. Ghosh,                     
\textit{``Cluster ensembles¡ªa knowledge reuse framework for
combining multiple partitions"}, The Journal of Machine
Learning Research 3 (2003) 583-617. 



\bibitem{29}                        
A. McDaid, D. Greene, and N. Hurley,                     
\textit{``Normalized Mutual Information to evaluate overlapping
community finding algorithms"},
CoRR (2011) abs/1110.2515  



\bibitem{30}                        
A. Lancichinetti, S. Fortunato, F. Radicchi,                        
\textit{``Benchmark graphs for testing community detection
algorithms."},
Phys. Rev. E 78 (2008) 046110.     




\bibitem{31}                        
W. W. Zachary,                        
\textit{``An information flow model for conflict and fission in
small groups"},
J. Anthro. Res. (1977) 452-473.     



\bibitem{32}                        
D. Lusseau,                        
\textit{``The emergent properties of dolphin social network"},
Proc. Biol. Sci. 270 (2003) 186-188.      


\bibitem{33}                        
Lada A. Adamic, Natalie Glance,                        
\textit{``The political blogosphere and the 2004 U.S. election:
divided they blog"},
LinkKDD '05, pp. 36-43.      


\bibitem{34}                        
M. E. J. Newman,                        
\textit{``The structure of scientific collaboration networks"},
Proc. Natl. Acad. Sci. USA 98 (2001) 404-409.    


\bibitem{35}                        
J. Leskovec, K. Lang, A. Dasgupta, M. Mahoney,                      
\textit{``Community Structure in Large Networks: Natural Cluster
Sizes and the Absence of Large Well-Defined Clusters"},
Internet Mathematics, 6 (2009) 29-123.    


\bibitem{36}                        
B. Viswanath,                        
\textit{``On the Evolution of User Interaction in Facebook"},
WOSN'09, pp. 37-42.     


\bibitem{37}                        
J. McAuley, J. Leskovec,                        
\textit{``Image Labeling on a Network: Using Social-Network Meta
data for Image Classification"},
ECCV¡¯14, pp. 828-841.     



\bibitem{38}                        
M. Y. Zhou, Z. Zhuo, S. M. Cai, Z. Q. Fu,                        
\textit{``Community structure revealed by phase locking"},
Chaos 24 (2014) 033128.    


\end{thebibliography}
\end{document}